\documentclass[11pt]{article}
\usepackage{amsmath}
\usepackage{amssymb}
\usepackage{bbm}
\usepackage{epsfig}
\allowdisplaybreaks[4]
\begin{document}
\title{One-loop stau masses in the effective potential approach}
\author{P.M. Ferreira \\ Dublin Institute for Advanced Studies, \\
Ireland}
\date{June, 2001} 
\maketitle
\noindent
{\bf Abstract.} We calculate the one-loop contributions to the tau slepton 
masses in the Minimal Supersymmetric Standard Model in the effective potential 
approach. For the majority of parameter space under study, those corrections 
are shown to elevate the value of the lightest stau mass.
\vspace{-9cm}
\begin{flushright}
DIAS-STP-01-14
\end{flushright}
\vspace{10cm}

Recently a full one-loop charge breaking effective potential was calculated 
for the case where, other than $H_1^0$ and $H_2^0$, the fields $\tau_L$ and
$\tau_R$ also acquire non-zero vevs~\cite{ccb}, $l$ and $\tau$ respectively. 
Its minimisation, and impact thereof on CCB bounds, was undertaken in 
ref.~\cite{min}. In this article we will not worry about CCB, instead simply 
note that the full dependence on $l$ and $\tau$ in the effective potential 
enables us to calculate the one-loop stau masses - they are approximated by the
second derivatives of the effective potential~\cite{erz}, a procedure shown to 
give very accurate results~\cite{brig}, at least for the Higgs masses. Recently 
this approach was used to calculate two-loop corrections to the CP-even Higgs 
boson masses~\cite{ram}. Following the conventions and notation of 
ref.~\cite{ccb}, we recall that we consider the MSSM with Yukawa couplings set 
to zero for the first and second generations, the superpotential of the model 
being given by
\begin{eqnarray}
W & = & \lambda_t H_2\,Q\,t_R \; + \; \lambda_b H_1\,Q\,b_R \; +\; \lambda_\tau
H_1\,L\,\tau_R \; +\; \mu H_2\,H_1 \;\;\; .
\end{eqnarray}
Supersymmetry is softly broken by adding to the potential explicit mass terms 
for the gauginos and scalar partners, and bilinear and trilinear terms similar
in form to those present in the superpotential above but multiplied by 
coefficients $B$ and $A_i$. When the fields $H_1^0$, $H_2^0$, $\tau_L$ and 
$\tau_R$ have vevs $v_1/\sqrt{2}$, $v_2/\sqrt{2}$, $l/\sqrt{2}$ and 
$\tau/\sqrt{2}$ respectively, the tree-level potential is given by
\begin{eqnarray}
V_0 & =& \frac{\lambda_\tau^2}{4} [v_1^2\,(l^2+\tau^2) + l^2\,\tau^2] \;-\;
\frac{\lambda_\tau}{\sqrt{2}}\,(A_\tau\,v_1 + \mu\,v_2)\,l\,\tau \;+ \;
\frac{1}{2} \,( m_1^2\,v_1^2 \; + \; m_2^2\,v_2^2\; + \; m_L^2\,l^2 \; +
\nonumber \\
 & & m_\tau^2\,\tau^2 )\;-\;B\,\mu\,v_1\, v_2 \;+\; \frac{g^{\prime 2}}{32}\,
(v_2^2-v_1^2  
-l^2+2\,\tau^2)^2 \; +\; \frac{g_2^2}{32}\,(v_2^2-v_1^2+l^2)^2\;\;\; . 
\label{eq:vc}
\end{eqnarray}
The one-loop contributions to the effective potential are given, as usual, by
\begin{equation}
\Delta V_1 \, =\, \sum_\alpha \, \frac{n_\alpha}{64\pi^2}\,M_\alpha^4\, \left(\,
\log \frac{M_\alpha^2}{M^2}\, - \, \frac{3}{2}\,\right) \;\;\; ,
\label{eq:v1cor}
\end{equation}
where the $M_\alpha$ are the tree-level masses of particles of spin $s_\alpha$, 
$M$ is the renormalisation scale and $n_\alpha = (-1)^{2 s_\alpha} (2 s_\alpha +
1) C_\alpha Q_\alpha$, with $C_\alpha$ the number of colour degrees of freedom
and $Q_\alpha$ counting the number of particle-antiparticle states. The presence
in the potential of vevs carrying electric charge complicates the mass matrices
considerably, by causing mixing between neutral and charged fields. For example,
the tau sneutrinos are mixed with the charged higgs fields, originating a three 
by three mass matrix~\cite{ccb}. 

\section{The effective potential approach}
The one-loop contributions to the stau masses in the effective potential 
approach (e.p.a.) will be given by the second derivatives of $\Delta V_1$ with 
respect to $l$ and $\tau$, so that we have
\begin{equation}
[m^2_{\tilde{\tau}}]_{ij}\; \simeq \; \frac{\partial^2 V_0}{\partial v_i 
\partial v_j} \,+\, \left. \sum_\alpha \, \frac{n_\alpha}{32\pi^2}\,M_\alpha^2\,
\frac{\partial^2 M_\alpha^2}{\partial v_i \partial v_j} \,\left(\,\log 
\frac{M_\alpha^2}{M^2}\, - \, 1\,\right) \right|_{l=\tau=0} \;\;\; ,
\label{eq:1l}
\end{equation}
with $\{v_i , v_j\} = \{l , \tau\}$, and we have used the fact that
\begin{equation}
\left. \frac{\partial M_\alpha^2}{\partial l}\right|_{l=\tau=0} \,=\,
\left.\frac{\partial M_\alpha^2}{\partial\tau}\right|_{l=\tau=0} \,=\, 0\;\;\; ,
\label{eq:der}
\end{equation}
as follows trivially from analysing the $\{l , \tau\}$ dependence of the 
tree-level potential~\eqref{eq:vc}. With mass matrices as large as six by six, 
it would seem impossible to find analytical expressions for~\eqref{eq:1l}. We 
can however apply the same trick that allowed us to compute the derivatives 
$\partial M_\alpha^2/\partial v_i$ in ref.~\cite{min}, to wit: the squared 
masses $M_\alpha^2$ are the solutions $\lambda$ of the eigenvalue equation $F\,=
\,\det ([M_\alpha^2]\,-\,\lambda\,\mathbbm{1})\,=\, 0$. This equation - a simple
polynomial of degree $n$ for an $n \times n$ mass matrix - implicitly determines
the $\lambda$ in terms of the vevs present in the theory. Using~\eqref{eq:der} 
as well as the implicit function theorem we obtain
\begin{equation}
\frac{\partial^2 M_\alpha^2}{\partial v_i \partial v_j}\, = \, -\, \left. 
\frac{\displaystyle{\frac{\partial^2 F}{\partial v_i \partial v_j}}}{
\displaystyle{\frac{ \partial F}{\partial \lambda}}}\,\right|_{\lambda = 
M_\alpha^2} \;\;\; .
\end{equation}
The derivatives of the determinant $F$ are very easy to compute - taking into 
account that $F$ is of the form 
\begin{equation}
F\,=\,\sum_{i, j, k, \ldots} a_{1i}a_{2j}a_{3k}\,\ldots \epsilon_{ijk\ldots}
\;\;\; ,
\end{equation}
we get 
\begin{equation}
\frac{\partial F}{\partial x} \,=\, \sum_{i, j, k, \ldots} \left( \frac{
\partial a_{1i}}{\partial x} a_{2j}a_{3k}\,\ldots \;+\; a_{1i}\frac{\partial 
a_{2j}}{\partial x} a_{3k}\,\ldots \;+\; a_{1i} a_{2j}\frac{\partial a_{3k}
}{\partial x}\,\ldots \;+\; \ldots\right)\,\epsilon_{ijk\ldots} \;\;\; ,
\end{equation}
which is to say, the derivative of the determinant of an $n \times n$ matrix 
becomes the sum of the determinants of $n$ matrices, each identical to the 
initial matrix except for a line replaced with derivatives of the original 
coefficients. Along similar lines we see that the second derivatives of $F$ 
would originate $n^2$ determinants. The final expressions obtained in this way
for~\eqref{eq:1l} are quite involved and we present them in the next section as 
a function of the sparticles' masses - these can be computed either analytically
(the only masses that are more involved are the neutralino's; see, for instance,
ref.~\cite{bbo}) or numerically. At this point we must recall that the e.p.a.
gives only an approximation to the physical masses - in fact, from general
principles one sees that~\cite{erz,brig}
\begin{equation}
[m^2_{\tilde{\tau}}]_{ij}^{e.p.a.}\; = \; \frac{\partial^2 V^{eff}}{\partial v_i
\partial v_j}\; = \; -\,\Gamma_{ij}(p^2=0) \;\;\; ,
\label{eq:epa}
\end{equation}
where the $\Gamma_{ij}(p^2)$ are the inverse propagators of the stau sector, 
given by~\footnote{We follow the conventions of ref.~\cite{brig} and absorb a 
factor of $i$ in the definition of $\Gamma$; furthermore, the $\hat{\Pi}$ are
understood as being the real part of the one-loop renormalised self-energies.}
\begin{equation}
\Gamma_{ij}(p^2) \; = \; p^2\,\delta_{ij} \,-\, M^2_{ij} \,+\,\hat{\Pi}_{ij}
(p^2) \;\;\; .
\label{eq:ga}
\end{equation}
The matrix $M^2_{ij}$ stands for the coefficients of the terms in the effective
potential which are quadratic in the stau fields, and is given by
\begin{equation}
M^2_{ij} \; = \; \begin{pmatrix} m_L^2 \,+\, \frac{1}{2}\,\left[2 (m_W^2)^\prime
- (m_Z^2)^\prime\right]\,\cos(2\beta) & \bar{m}_\tau^\prime\,(A_\tau\,-\,\mu
\cot \beta) \vspace{0.2cm} \\
\bar{m}_\tau^\prime\,(A_\tau \,-\,\mu \cot \beta) & m_\tau^2 \,+\, \left[
(m_W^2)^\prime - (m_Z^2)^\prime \right]\,\cos(2\beta) \end{pmatrix} \;\;\; ,
\label{eq:mij}
\end{equation}
where $\bar{m}_\tau$ represents the tau fermion mass. The parameters in this 
mass matrix are all one-loop renormalised. The primes in the masses indicate 
that these are the masses in the e.p.a. approximation, that is, $(m_W^2)^\prime 
= g_2^2 (v_1^2+ v_2^2)/4$, $(m_Z^2)^\prime = (g^{\prime 2}+g_2^2) (v_1^2+
v_2^2)/4$ and $\bar{m}_\tau^\prime = \lambda_\tau v_1/\sqrt{2}$. These are 
related to the physical, ``unprimed" masses by
\begin{eqnarray}
(m_W^2)^\prime & = & m_W^2 \;+\; \hat{\Pi}_{WW}(m_W^2) \;\;\; , \nonumber \\
(m_Z^2)^\prime & = & m_Z^2 \;+\; \hat{\Pi}_{ZZ}(m_Z^2) \;\;\; , \nonumber \\
(m_\tau)^\prime & = & m_\tau \;+\; \hat{\Pi}_{\tau\tau}(m_\tau) \;\;\; . 
\end{eqnarray}
Our input parameters are the experimental values of the masses, so we express
$M^2_{ij}$ in terms of unprimed quantities, so that $M^2_{ij} \,=\, 
\bar{M}^2_{ij} \,+\, \bar{\Pi}_{ij}$ - the matrix $\bar{M}^2_{ij}$ is identical 
in form to eq.~\eqref{eq:mij}, except the ``primed" masses are replaced by the 
physical ones, and we have
\begin{equation}
\bar{\Pi}_{ij} \; = \;\begin{pmatrix} \frac{1}{2}\,\left[2 \hat{\Pi}_{WW}(m_W^2)
- \hat{\Pi}_{ZZ}(m_Z^2) \right]\,\cos(2\beta) & \hat{\Pi}_{\tau\tau}(m_\tau)\,
(A_\tau\,-\,\mu \cot \beta) \vspace{0.2cm} \\   
\hat{\Pi}_{\tau\tau}(m_\tau)\,(A_\tau \,-\,\mu \cot\beta) & 
\hat{\Pi}_{\tau\tau}(m_\tau) \,+\, \left[\hat{\Pi}_{WW}(m_W^2) - 
\hat{\Pi}_{ZZ}(m_Z^2)\right]\,\cos(2\beta) 
\end{pmatrix} \;\;\; .
\end{equation}
So, we can rewrite the inverse propagators of eq.~\eqref{eq:ga} as
\begin{equation}
\Gamma_{ij}(p^2) \; = \; p^2\,\delta_{ij} \,-\, \bar{M}^2_{ij} \,+\,
\hat{\Pi}_{ij} (0) \,+\, \Delta_{ij} (p^2) \;\;\;,
\end{equation}
where we define $\Delta_{ij} (p^2)$ as
\begin{equation}
\Delta_{ij} (p^2) \; = \; \hat{\Pi}_{ij} (p^2) \,-\,\bar{\Pi}_{ij} \,-\,
\hat{\Pi}_{ij} (0) \;\;\; .
\end{equation}
Thus, the e.p.a. expressed by eq.~\eqref{eq:epa} consists of neglecting the 
quantity $\Delta_{ij} (p^2)$ - which is equivalent to calculating the stau 
self energies $\hat{\Pi}_{ij}$ at zero external momentum and neglecting the
effects of the gauge boson and tau self energies present in $\bar{\Pi}_{ij}$. 
The careful comparison between the diagrammatic and the effective potential 
approaches of ref.~\cite{brig} reached similar conclusions for the higgs' 
masses. It was shown there that the e.p.a. produces extremely good 
results, the masses thus calculated differing from the ``real" masses by very 
small amounts, at most $\sim$ 3 GeV. It is reasonable to expect, then, that 
the e.p.a.-calculated stau masses will be comparably accurate. We can also 
expect the one-loop mass corrections to be small - the second derivatives of the
masses with order to $\{l , \tau\}$ will always produce coefficients multiplied
by the smaller couplings, $g^\prime$, $g_2$ and $\lambda_\tau$.

\section{One-loop stau masses}
We now list all one-loop contributions to the stau mass matrix as per 
eq.~\eqref{eq:1l}. The sparticle masses are well known in the literature and
the factors $n_\alpha$ are given in ref.~\cite{ccb}~\footnote{We take this 
opportunity to correct a misprint in ref.~\cite{ccb}, though: the factor 
$n_\alpha$ for the first and second generation sneutrinos is 4, counting two 
generations and the existence of both neutrinos and anti-neutrinos.}, so we list
only the non-zero second derivatives of the masses. We adopt the convention 
$X_{v_i v_j}$ to denote the second derivative of the quantity $X$ with respect 
to $\{v_i , v_j\} \,=\,\{l , \tau\}$ (and recall that these derivatives are 
evaluated at $l = \tau=0$). The simplest of these contributions are those of the
first and second generation squarks and sleptons and the gauge bosons, given by
\begin{align}
M^2_{\tilde{u}_1,ll} & = -\frac{1}{12}({g^\prime}^2+3\,g_2^2) &
M^2_{\tilde{u}_1,\tau\tau} & = \frac{1}{6}\,{g^\prime}^2 &
M^2_{\tilde{u}_2,ll} & = \frac{1}{3}\,{g^\prime}^2 \\
M^2_{\tilde{u}_2,\tau\tau} & = -\frac{2}{3}\,{g^\prime}^2 &
M^2_{\tilde{d}_1,ll} & =  \frac{1}{12}(3\,g_2^2-{g^\prime}^2) &
M^2_{\tilde{d}_1,\tau\tau} & = \frac{1}{6}\,{g^\prime}^2 \nonumber \\
M^2_{\tilde{d}_2,ll} & = -\frac{1}{6}\,{g^\prime}^2 &
M^2_{\tilde{d}_2,\tau\tau} & = \frac{1}{3}\,{g^\prime}^2 &
M^2_{\tilde{e}_1,ll} & = \frac{1}{4}(g_2^2+{g^\prime}^2) \nonumber \\
M^2_{\tilde{e}_1,\tau\tau} & = -\frac{1}{2}{g^\prime}^2 &
M^2_{\tilde{e}_2,ll} & = -\frac{1}{2}{g^\prime}^2 &
M^2_{\tilde{e}_2,\tau\tau} & = {g^\prime}^2 \nonumber \\
M^2_{\tilde{\nu}_e,ll} & = \frac{1}{4}({g^\prime}^2-g_2^2) &
M^2_{\tilde{\nu}_e,\tau\tau} & = -\frac{1}{2}{g^\prime}^2 &
M^2_{W,ll} & = \frac{1}{2}\, g_2^2 \nonumber \\
M^2_{Z,ll} & = \frac{1}{2}\, (g_2^2+{g^\prime}^2)\,\cos^2(2\theta_W) 
\end{align}
For the stop, we have
\begin{eqnarray}
M^2_{\tilde{t}_1,ll} & = & \frac{3\,({g^\prime}^2-g_2^2)\,M^2_{\tilde{t}_1} \,-
\,4\,{g^\prime}^2\,a_{\tilde{t}}\,+\,({g^\prime}^2+3g_2^2)\,c_{\tilde{t}}}{12\,
\left(M^2_{\tilde{t}_1}\,-\, M^2_{\tilde{t}_2}\right)} \nonumber \\
M^2_{\tilde{t}_1,\tau\tau} & = & -\,\frac{{g^\prime}^2}{6}\,\frac{3\,
M^2_{\tilde{t}_1}\,-\,4 \,a_{\tilde{t}}\,+\,c_{\tilde{t}}}{M^2_{\tilde{t}_1}\,-
\, M^2_{\tilde{t}_2}} \;\;\; ,
\end{eqnarray}
with 
\begin{eqnarray}
a_{\tilde{t}} & = & m_Q^2 \;+\;\frac{1}{2}\,\lambda_t^2\,v_2^2 \;+\;\frac{1}{24}
({g^\prime}^2\,-\,3\,g_2^2)\,(v_2^2-v_1^2) \nonumber \\
c_{\tilde{t}} & = & m_t^2 \;+\;\frac{1}{2}\,\lambda_t^2\,v_2^2 \;-\;\frac{
{g^\prime}^2}{6}\,(v_2^2-v_1^2) \;\;\; ,
\end{eqnarray}
and identical expressions for the second stop, with the substitution 
$M^2_{\tilde{t}_1} \leftrightarrow M^2_{\tilde{t}_2}$. For the sbottom we have
\begin{eqnarray}
M^2_{\tilde{b}_1,ll} & = & \frac{3\,(g_2^2-{g^\prime}^2)\,M^2_{\tilde{b}_1} \,+
\,2\,{g^\prime}^2\,a_{\tilde{b}}\,+\,({g^\prime}^2-3g_2^2)\,c_{\tilde{b}}}{12\,
\left(M^2_{\tilde{b}_1}\,-\, M^2_{\tilde{b}_2}\right)} \nonumber \\
M^2_{\tilde{b}_1,\tau\tau} & = & \frac{{g^\prime}^2}{6}\,\frac{3\,
M^2_{\tilde{b}_1}\,-\,2\,a_{\tilde{b}}\,-\,c_{\tilde{b}}}{M^2_{\tilde{b}_1}\,-\,
M^2_{\tilde{b}_2}} \;\;\; ,
\end{eqnarray}
with
\begin{eqnarray}
a_{\tilde{b}} & = & m_Q^2 \;+\;\frac{1}{2}\,\lambda_b^2\,v_1^2 \;+\;\frac{1}{24}
({g^\prime}^2\,+\,3\,g_2^2)\,(v_2^2-v_1^2) \nonumber \\
c_{\tilde{b}} & = & m_b^2 \;+\;\frac{1}{2}\,\lambda_b^2\,v_1^2 \;+\;\frac{
{g^\prime}^2}{12}\,(v_2^2-v_1^2) \;\;\; .
\end{eqnarray}
Again, identical expressions for the second sbottom with trivial substitutions. 
For the charginos we find
\begin{align}
M^2_{\chi^\pm_1,ll} & = g_2^2\;\frac{\mu^2\,+\,\frac{1}{2}\,g_2^2\,v_1^2\;-\;
M^2_{\chi^\pm_1}}{M^2_{\chi^\pm_2}\,-\,M^2_{\chi^\pm_1}} & 
M^2_{\chi^\pm_1,\tau\tau} & = \lambda_\tau^2\;\frac{M_2^2\,+\,\frac{1}{2}\,
g_2^2\, v_2^2\,-\,M^2_{\chi^\pm_1}}{M^2_{\chi^\pm_2}\,-\,M^2_{\chi^\pm_1}}  
\nonumber \\
M^2_{\chi^\pm_1,l\tau} & = \frac{\lambda_\tau\, g_2^2}{\sqrt{2}}\;\frac{\mu
\,v_2\, +\, M_2\,v_1}{M^2_{\chi^\pm_2}\,-\,M^2_{\chi^\pm_1}} \;\;\; . & & 
\end{align}
For the neutralinos and tau lepton the expressions are more complex. From the
formulae of ref.~\cite{min} it is easy to find~\footnote{Notice that this is
the second derivative of the mass, not its squared.} 
\begin{equation}
M_{\chi^0_i,xy} \;=\;-\, \frac{-B_{\chi^0,xy}\,M_{\chi^0_i}^4 - C_{\chi^0,xy}\,
M_{\chi^0_i}^3 + D_{\chi^0,xy}\,M_{\chi^0_i}^2 + E_{\chi^0,xy}\,M_{\chi^0_i} + 
F_{\chi^0,xy}}{6\,M_{\chi^0_i}^5 - 5\,A_{\chi^0}\,M_{\chi^0_i}^4 - 4\,B_{\chi^0}
\,M_{\chi^0_i}^3 - 3\,C_{\chi^0}\,M_{\chi^0_i}^2 + 2\,D_{\chi^0}\,M_{\chi^0_i} +
E_{\chi^0}} \;\;\; ,
\end{equation}
where the index $i$ runs from 1 to 6, the two last entries being $\pm 
\lambda_\tau\,v_1/\sqrt{2}$ (the tau mass) and the coefficients $A_{\chi^0}, 
\ldots F_{\chi^0}$ are given by
\begin{eqnarray}
A_{\chi^0} & = & M_1\,+\,M_2 \nonumber \\
B_{\chi^0} & = & \frac{\lambda_\tau^2}{2}\,v_1^2\;+\;\frac{1}{4}\,({g^\prime}^2
\,+\,g_2^2)\,(v_1^2\,+\,v_2^2)\;+ \;\mu^2\;-\;M_1\,M_2 \nonumber \\
C_{\chi^0} & = & -\,\frac{\lambda_\tau^2}{2}\,(M_1\,+M_2)\,v_1^2\;-\;\frac{1}{4}
\,({g^\prime}^2\,M_2\,+\,g_2^2\,M_1)\,(v_1^2\,+\,v_2^2)\;+\;\frac{1}{2}\,
({g^\prime}^2\,+\,g_2^2)\,\mu\,v_1\,v_2\;-\nonumber \\
 & & \mu^2\,(M_1\,+\,M_2) \nonumber \\
D_{\chi^0} & = & \lambda_\tau^2\,v_1^2\, \left[ \frac{1}{8}\,({g^\prime}^2\,+\,
g_2^2)\,(v_1^2\,+\,v_2^2)\;-\;\frac{1}{2}\,M_1\,M_2\;+\;\frac{\mu^2}{2} \right]
\;+\;\frac{1}{2}\,({g^\prime}^2\,M_2\,+\,g_2^2\,M_1)\,\mu\,v_1\,v_2\;-
\nonumber \\
 & & M_1\, M_2\,\mu^2 \nonumber \\
E_{\chi^0} & = & \lambda_\tau^2\,v_1^2\, \left[ \frac{1}{4}\,({g^\prime}^2\,+\,
g_2^2)\,\mu\,v_1\,v_2\,-\,\frac{1}{8}\, ({g^\prime}^2\,M_2\,+\,g_2^2\,M_1)\,
(v_2^2\,-\,2\,v_1^2) \,-\, \frac{\mu^2}{2}\,(M_1\,+\,M_2)\right] 
\end{eqnarray}
and their derivatives,
\begin{eqnarray}
B_{\chi^0,ll} & = & \frac{1}{2}\,(2\,\lambda_\tau^2\,+\,{g^\prime}^2\,+\,g_2^2)
\hspace{1.7cm} B_{\chi^0,\tau} \; = \; (\lambda_\tau^2\,+\,2\,{g^\prime}^2)
\nonumber \\
C_{\chi^0,ll} & = & -\,\frac{1}{2}\,\left[2\,\lambda_\tau^2\,(M_1\,+\,M_2)\;+\;(
{g^\prime}^2\,M_2\,+\,g_2^2\,M_1)\right]\nonumber \\
C_{\chi^0,\tau\tau} & = & -\,\left[\lambda_\tau^2\,(M_1\,+\,M_2)\;+\;2\,
{g^\prime}^2 \,M_2\right] \nonumber \\
C_{\chi^0,l\tau} & = & -\,\frac{\lambda_\tau}{2\sqrt{2}}\,(3\,{g^\prime}^2\,-\,
g_2^2\,-\,2\, \lambda_\tau^2)\,v_1 \nonumber \\
D_{\chi^0,ll} & = & \lambda_\tau^2\,\left[\frac{1}{4}\,({g^\prime}^2\,+\,g_2^2)
\,(v_2^2\,-\,2\,v_1^2) \;-\;M_1\,M_2\right] \;+\;\frac{\mu^2}{2}\,({g^\prime}^2
\,+\,g_2^2) \nonumber \\
D_{\chi^0,\tau\tau} & = & \lambda_\tau^2\,\left[{g^\prime}^2\,v_1^2\,+\,\frac{1}{4}\,({g^\prime}^2\,+\,g_2^2)\,v_2^2 - M_1\,M_2\right]\;+\;\frac{1}{2}\,\left[
{g^\prime}^2\,g_2^2\,(v_1^2\,+\,v_2^2)\,+\,4\,{g^\prime}^2\,\mu^2\right] \nonumber \\
D_{\chi^0,l\tau} & = & \frac{\lambda_\tau^3}{\sqrt{2}}\,(M_1+M_2)\,v_1\;+\;
\frac{\lambda_\tau}{2\sqrt{2}}\,\left[(g_2^2\,M_1\,-\,3\,{g^\prime}^2\,M_2)
\,v_1\,+\,({g^\prime}^2\,-\,g_2^2)\,\mu\,v_2\right] \nonumber \\
E_{\chi^0,ll} & = & \frac{\lambda_\tau^2}{4}\,\left[({g^\prime}^2\,M_2\,+\,g_2^2
\,M_1)\,(2\,v_1^2\,-\,v_2^2)\;-\;2\,({g^\prime}^2\,+\,g_2^2)\,\mu\,v_1\,v_2
\right]\;-\;\frac{\mu^2}{2}\,({g^\prime}^2\,M_2\,+\,g_2^2\,M_1) \nonumber \\
E_{\chi^0,\tau\tau} & = & \frac{\lambda_\tau^2}{4}\,\left[4\,{g^\prime}^2\,(\mu
\,v_1\,v_2\,-\,M_2\,v_1^2)\,-\,({g^\prime}^2\,M_2\,+\,g_2^2\,M_1)\,v_2^2\right] 
\;+\; {g^\prime}^2\,(g_2^2\,v_1\,v_2\,-\,2\,\mu\,M_2)\,\mu \nonumber \\
E_{\chi^0,l\tau} & = & \frac{\lambda_\tau^3}{4\sqrt{2}}\,\left[({g^\prime}^2\,+
\, g_2^2)\,v_2^2\;-\;4\,M_1\,M_2\right]\,v_1 \;+\nonumber \\
 & & \frac{\lambda_\tau}{2\sqrt{2}}\,\left[\mu\,(g_2^2\,M_1\,-\,{g^\prime}^2\,
M_2)\,v_2\;-\;2\,{g^\prime}^2\,\mu^2\,v_1\right] \nonumber \\
F_{\chi^0,ll} & = & \frac{\lambda_\tau^2}{2}\,({g^\prime}^2\,M_2\,+\,g_2^2\,M_1)
\,\mu\,v_1\,v_2 \nonumber \\
F_{\chi^0,\tau\tau} & = & -\,\lambda_\tau^2\,{g^\prime}^2\,\mu\,M_2\,v_1\,
v_2 \nonumber \\
F_{\chi^0,l\tau} & = & -\frac{\lambda_\tau^3}{4\sqrt{2}}\,({g^\prime}^2\,M_2\,+
\, g_2^2\,M_1)\,v_1\,v_2^2\;+\;\frac{\lambda_\tau}{\sqrt{2}}\,{g^\prime}^2\,
\mu^2\,M_2\,v_1\;\;\;.
\end{eqnarray}
For the charged higgses, 
\begin{eqnarray}
M^2_{H^\pm_1,ll} & = &-\, \frac{(2\lambda_\tau^2-g_2^2)^2
\,v_1^2\,\bigl(c_\pm-M^2_{H^\pm_1}\bigr)+g_2^2\,v_2\left[g_2^2\,v_2\bigl(a_\pm-
M^2_{H^\pm_1}\bigr)+2 (2\lambda_\tau^2-g_2^2)v_1b_\pm\right]}{8\bigl(f_\pm
-M^2_{H^\pm_1}\bigr)\bigl(a_\pm+c_\pm-2M^2_{H^\pm_1}\bigr)} \; +\nonumber \\
 & & \frac{1}{4}\,({g^\prime}^2+g_2^2)\,\frac{c_\pm-a_\pm}{a_\pm+c_\pm-2 
M^2_{H^\pm_1}} \nonumber \\
M^2_{H^\pm_1,\tau\tau} & = & -\,\lambda_\tau^2\,\frac{A_\tau^2\bigl(c_\pm-
M^2_{H^\pm_1}\bigr)+\mu\left[\mu\bigl(a_\pm-M^2_{H^\pm_1}\bigr)-2 A_\tau b_\pm
\right]}{\bigl(f_\pm-M^2_{H^\pm_1}\bigr)\bigl(a_\pm+c_\pm-2M^2_{H^\pm_1}\bigr)} 
\;+ \nonumber \\
 & & \frac{{g^\prime}^2\,(a_\pm-c_\pm)\,+\,2\lambda_\tau^2\bigl(c_\pm-
M^2_{H^\pm_1}\bigr)}{2\,\bigl(a_\pm+c_\pm-2M^2_{H^\pm_1}\bigr)} \nonumber \\
M^2_{H^\pm_1,l\tau} & = & \lambda_\tau\,\frac{(2\lambda_\tau^2-g_2^2)v_1
\left[\mu b_\pm - A_\tau\bigl(c_\pm-M^2_{H^\pm_1}\bigr)\right]+g_2^2\,v_2
\left[\mu\bigl(a_\pm-M^2_{H^\pm_1}\bigr)-A_\tau b_\pm\right]}{2\sqrt{2}\bigl(
f_\pm- M^2_{H^\pm_1}\bigr)\bigl(a_\pm+c_\pm-2M^2_{H^\pm_1}\bigr)} 
\end{eqnarray}
with coefficients
\begin{align}
a_\pm & = m_1^2\;-\; \frac{{g^\prime}^2}{8}\,(v_2^2-v_1^2)\;+\;\frac{g_2^2}{8}
\,(v_2^2+v_1^2) &
b_\pm & = B\,\mu\; + \; \frac{g_2^2}{4}\,v_1\,v_2 \nonumber \\
c_\pm & = m_2^2 \;+\; \frac{{g^\prime}^2}{8}\,(v_2^2-v_1^2)\;+\;
\frac{g_2^2}{8}\,(v_2^2+v_1^2) &
f_\pm & = m_L^2\;-\;\frac{1}{8}\,({g^\prime}^2+g_2^2)\,(v_2^2-v_1^2) \;\;\; .
\end{align}
Similar expressions hold for $M^2_{H^\pm_2}$ - notice that if we perform a 
tree-level minimisation of the potential, this second eigenvalue is actually 
zero, reflecting the presence of a Goldstone boson. With a one-loop minimisation
the tree-level charged higgs mass matrix produces two non-zero eigenvalues,
albeit the second is quite small when compared to the first one, and therefore
has a small impact in the one-loop potential. This second eigenvalue can even be
negative~\cite{gam} and its contribution to the potential neglected following 
the arguments of~\cite{rai}. For the tau sneutrino we have
\begin{eqnarray}
M^2_{\tilde{\nu}_\tau,ll} & = & -\,\frac{(2\lambda_\tau^2-g_2^2)^2\,v_1^2\bigl(
c_\pm -M^2_{\tilde{\nu}_\tau}\bigr)+g_2^2\,v_2\left[g_2^2\,v_2\bigl(a_\pm-
M^2_{\tilde{\nu}_\tau}\bigr)+2 (2\lambda_\tau^2-g_2^2)v_1 b_\pm\right]}{8\bigl(
a_\pm-M^2_{\tilde{\nu}_\tau}\bigr)\bigl(c_\pm-M^2_{\tilde{\nu}_\tau}\bigr)-8 
b_\pm^2} \;+ \nonumber \\
 & & \frac{1}{4}({g^\prime}^2+g_2^2) \nonumber \\
M^2_{\tilde{\nu}_\tau,\tau\tau} & = & \lambda_\tau^2\,-\,\frac{1}{2}\,
{g^\prime}^2\;-\;\lambda_\tau^2\,\frac{A_\tau^2\bigl(c_\pm -
M^2_{\tilde{\nu}_\tau}\bigr)+\mu\left[\mu \bigl(a_\pm-M^2_{\tilde{\nu}_\tau}
\bigr)-2 A_\tau b_\pm\right]}{\bigl(a_\pm - M^2_{\tilde{\nu}_\tau}\bigr)\bigl(
c_\pm- M^2_{\tilde{\nu}_\tau}\bigr)- b_\pm^2} \nonumber \\
M^2_{\tilde{\nu}_\tau,l\tau} & = & \lambda_\tau\,\frac{(2\lambda_\tau^2-g_2^2)
v_1 \left[\mu b_\pm - A_\tau\bigl(c_\pm-M^2_{\tilde{\nu}_\tau}\bigr)\right]+
g_2^2\,v_2\left[\mu\bigl(a_\pm-M^2_{\tilde{\nu}_\tau}\bigr)-A_\tau b_\pm\right]}
{2\sqrt{2}\left[\bigl(a_\pm - M^2_{\tilde{\nu}_\tau}\bigr)\bigl(c_\pm- 
M^2_{\tilde{\nu}_\tau} \bigr)- b_\pm^2\right]} \;\;\; .
\end{eqnarray}
The reason for the sharing of coefficients between the second derivatives of 
$M^2_{H^\pm}$ and $M^2_{\tilde{\nu}_\tau}$ is the mixing discussed in 
ref.~\cite{ccb}. For the pseudo-scalar Higgs the second derivatives are given by
\begin{eqnarray}
M^2_{\bar{H}_1,ll} & = & \frac{4\lambda_\tau^2\,\bigl( c_{\bar{H}}-
M^2_{\bar{H}_1}\bigr)\,+\,({g^\prime}^2-g_2^2)\,(c_{\bar{H}}-a_{\bar{H}})}{4 
\bigl( M^2_{\bar{H}_2}\,-\,M^2_{\bar{H}_1}\bigr) } \;+\;
\frac{A_{ll}(M^2_{\bar{H}_1})}{D_{\bar{H}}\bigl(M^2_{\bar{H}_1},M^2_{\bar{H}_2}
\bigr)} \nonumber \\
M^2_{\bar{H}_1,\tau\tau} & = & \frac{2\lambda_\tau^2\,\bigl( c_{\bar{H}}-
M^2_{\bar{H}_1}\bigr)\,-\,{g^\prime}^2\,(c_{\bar{H}}-a_{\bar{H}})}{2\bigl(
M^2_{\bar{H}_2}\,-\,M^2_{\bar{H}_1}\bigr)}\;+\; 
\frac{A_{\tau\tau}(M^2_{\bar{H}_1})}{D_{\bar{H}}\bigl(M^2_{\bar{H}_1},
M^2_{\bar{H}_2}\bigr)}\nonumber \\
M^2_{\bar{H}_1,l\tau} & = & \frac{A_{l\tau}(M^2_{\bar{H}_1})}{D_{\bar{H}}\bigl(
M^2_{\bar{H}_1},M^2_{\bar{H}_2}\bigr)}
\end{eqnarray}
where the functions $A$ are given by
\begin{eqnarray}
A_{ll}(M^2_{\bar{H}_1}) & = & \lambda_\tau^2\,\bigl( h_{\bar{H}}-M^2_{\bar{H}_1}
\bigr)\,\left[ 2\, B\,\mu^2\,A_\tau\,-\,\mu^2\,\bigl( a_{\bar{H}}-
M^2_{\bar{H}_1} \bigr)\,-\,A_\tau^2\,\bigl( c_{\bar{H}}-M^2_{\bar{H}_1}\bigr)
\right] 
\nonumber \\
A_{\tau\tau}(M^2_{\bar{H}_1}) & = & \lambda_\tau^2\,\bigl( j_{\bar{H}}- 
M^2_{\bar{H}_1}\bigr)\,\left[2\,B\,\mu^2\,A_\tau\,-\,\mu^2\,\bigl( a_{\bar{H}}- 
M^2_{\bar{H}_1}\bigr)\,-\, A_\tau^2\,\bigl( c_{\bar{H}}-M^2_{\bar{H}_1}\bigr)
\right] \nonumber \\
A_{l\tau}(M^2_{\bar{H}_1}) & = & \lambda_\tau^2\,i_{\bar{H}}\,\left[\mu^2\,
\bigl( a_{\bar{H}} -M^2_{\bar{H}_1}\bigr)\,+\,A_\tau^2\,\bigl( c_{\bar{H}}-
M^2_{\bar{H}_1}\bigr)\,- \,2\,B\,\mu^2\,A_\tau \right]
\end{eqnarray}
and the denominator $D_{\bar{H}}$ by
\begin{equation}
D_{\bar{H}}\bigl(M^2_{\bar{H}_1},M^2_{\bar{H}_2}\bigr) \; = \; \bigl( 
M^2_{\bar{H}_2}\,-\,M^2_{\bar{H}_1}\bigr)\,\left[\bigl( h_{\bar{H}}- 
M^2_{\bar{H}_1}\bigr)\,\bigl(j_{\bar{H}}-M^2_{\bar{H}_1}\bigr)\,-\,
{i_{\bar{H}}}^2\right]
\end{equation}
with coefficients
\begin{align}
a_{\bar{H}} & = m_1^2\;-\;\frac{1}{8}\,({g^\prime}^2 + g_2^2)\,(v_2^2-v_1^2) &
c_{\bar{H}} & = m_2^2\;+\;\frac{1}{8}\,({g^\prime}^2 + g_2^2)\,(v_2^2-v_1^2)
\nonumber \\
h_{\bar{H}} & = m_L^2\;+\; \frac{1}{8}\,(g_2^2 - {g^\prime}^2)\,(v_2^2-v_1^2) &
i_{\bar{H}} & = \frac{\lambda_\tau}{\sqrt{2}}\,(\mu\,v_2\;-\;A_\tau\,v_1) 
\nonumber \\
j_{\bar{H}} & = m_\tau^2\;+\;\frac{{g^\prime}^2}{4}\,(v_2^2-v_1^2) \;\;\; . & &
\end{align}
With the replacement $M^2_{\bar{H}_1}\leftrightarrow M^2_{\bar{H}_2}$, we obtain
the expressions for the second pseudo-scalar~\footnote{Which, like in the case 
of the charged Higgses, is a Goldstone boson for a tree-level minimisation.}. 
For the Higgs scalars, we obtain
\begin{eqnarray}
M^2_{h,ll} & = & \frac{4\,\lambda_\tau^2\,(c_H\,-\,M^2_h)\,+\,({g^\prime}^2-
g_2^2)\,(c_H\,-\,a_H)}{4\,\left(M^2_H\,-\,M^2_h\right)}\;+\;\frac{B_{ll}
(M^2_h)}{D_H(M^2_h,M^2_H)} \nonumber \\
M^2_{h,\tau\tau} & = & \frac{2\,\lambda_\tau^2\,(c_H\,-\,M^2_h)\,+\,{g^\prime}^2
\,(a_H\,-\,c_H)}{2\,\left(M^2_H\,-\,M^2_h\right)}\;+\;\frac{B_{\tau\tau}
(M^2_h)}{D_H(M^2_h,M^2_H)} \nonumber \\
M^2_{h,l\tau} & = & \frac{B_{l\tau}(M^2_h)}{D_H(M^2_h,M^2_H)}
\label{eq:mh1}
\end{eqnarray}
where the denominator $D_H$ has the expression 
\begin{equation}
D_H(M^2_h,M^2_H)\;=\;\left(M^2_H\,-\,M^2_h\right)\,\left[\left(h_{\bar{H}}\,-\,
M^2_h\right)\,\left(j_{\bar{H}}\,-\,M^2_h\right)\,-\,i_{\bar{H}}^2\right]
\label{eq:mh2}
\end{equation}
and the functions $B$ are given by
\begin{align}
B_{ll}(M^2_h) = \;& 2\,(a_H\,-\,M^2_h)\,\left[2\,i_H\,f_{H,l}\,g_{H,l}\,
-\,f_{H,l}^2\,(j_{\bar{H}}\,-\,M^2_h)\,-\,g_{H,l}^2\,(h_{\bar{H}}\,-\,M^2_h)
\right]\;+ \vspace{0.3cm} \nonumber \\
 & 2\,(c_H\,-\,M^2_h)\,\left[2\,i_H\,d_{H,l}\,e_{H,l}\,-\,d_{H,l}^2\,
(j_{\bar{H}}\,-\,M^2_h)\,-\,e_{H,l}^2\,(h_{\bar{H}}\,-\,M^2_h)\right]\;+ 
\vspace{0.3cm} \nonumber \\
 & 4\,b_H\left[d_{H,l}\,f_{H,l}\,(j_{\bar{H}}-M^2_h)\,+\,e_{H,l}\,g_{H,l}
\,(h_{\bar{H}}-M^2_h)\,-\,i_H (g_{H,l}\,d_{H,l}+e_{H,l}\,f_{H,l}) 
\right]\vspace{0.3cm} \nonumber \\
B_{\tau\tau}(M^2_h) = \;& 2\,(a_H\,-\,M^2_h)\,\left[2\,i_H\,f_{H,\tau}\,
g_{H,\tau}\,-\,f_{H,\tau}^2\,(j_{\bar{H}}\,-\,M^2_h)\,-\,g_{H,\tau}^2\,
(h_{\bar{H}}\,-\,M^2_h)\right] \;+ \vspace{0.3cm} \nonumber \\
 & 2\,(c_H\,-\,M^2_h)\,\left[2\,i_H\,d_{H,\tau}\,e_{H,\tau}\,-\,
d_{H,\tau}^2\,(j_{\bar{H}}\,-\,M^2_h)\,-\,e_{H,\tau}^2\,(h_{\bar{H}}\,-\,M^2_h)
\right]\;+\vspace{0.3cm} \nonumber \\
 &  4\, b_H \left[d_{H,\tau}\,f_{H,\tau}\,(j_{\bar{H}}-M^2_h)\,+\,e_{H,\tau}
\,g_{H,\tau}\,(h_{\bar{H}}-M^2_h)\,-\,i_H (g_{H,\tau}\,d_{H,\tau}+
e_{H,\tau}\,f_{H,\tau}) \right] \vspace{0.3cm} \nonumber \\
B_{l\tau}(M^2_h) = \;& 2\,(a_H-M^2_h)\left[i_H (f_{H,l}\,g_{H,\tau}+
f_{H,\tau}\,g_{H,l})-f_{H,l}\,f_{H,\tau}(j_{\bar{H}}-M^2_h)-g_{H,l}\,g_{H,\tau}
(h_{\bar{H}}-M^2_h)\right] \,+ \vspace{0.3cm} \nonumber \\
 & 2\,(c_H-M^2_h)\left[i_H (d_{H,l}\,e_{H,\tau}+d_{H,\tau}\,e_{H,l})-
d_{H,l}\,d_{H,\tau}(j_{\bar{H}}-M^2_h)-e_{H,l}\,e_{H,\tau}(h_{\bar{H}}-M^2_h)
\right] \,+ \vspace{0.3cm} \nonumber \\
 & 2\, b_H\left[(e_{H,l}\,g_{H,\tau}+e_{H,\tau}\,g_{H,l})(h_{\bar{H}}-M^2_h)+
(d_{H,l}\,f_{H,\tau}+d_{H,\tau}\,f_{H,l})(j_{\bar{H}}-M^2_h)\,- \right. 
\vspace{0.3cm}\nonumber \\
 & \left. i_H (e_{H,\tau}\,f_{H,l}+e_{H,l}\,f_{H,\tau}+d_{H,l}\,
g_{H,\tau}+ d_{H,\tau}\,g_{H,l}) \right] \;\;\;.
\label{eq:mh3}
\end{align}
The new coefficients in these expressions are
\begin{align}
a_H & = m_1^2\;-\;\frac{1}{8}\,({g^\prime}^2+g_2^2)\,(v_2^2-3\,v_1^2) &
b_H & = -\,B\,\mu \;-\;\frac{1}{4}\,({g^\prime}^2+g_2^2)\,v_1\,v_2 \nonumber \\
c_H & = m_2^2\;+\;\frac{1}{8}\,({g^\prime}^2+g_2^2)\,(3\,v_2^2-v_1^2) & 
i_H & = -\,i_{\bar{H}} 
\end{align}
and the derivatives of $\{e_H, d_H, f_H, g_H \}$ are listed in ref.~\cite{min}.
For completeness, they are
\begin{align}
d_{H,l} & = \frac{1}{4}(4\,\lambda_\tau^2\,+\,{g^\prime}^2-g_2^2)\,v_1 &
d_{H,\tau} & = \frac{\lambda_\tau}{\sqrt{2}}\,A_\tau &
e_{H,l} & = \frac{\lambda_\tau}{\sqrt{2}}\,A_\tau \nonumber \\
e_{H,\tau} & = \frac{1}{2}(2\,\lambda_\tau^2\,-\,{g^\prime}^2)\,v_1 &
f_{H,l} & = -\frac{1}{4}\,({g^\prime}^2\,-\,g_2^2)\,v_2&
f_{H,\tau} & = -\frac{\lambda_\tau}{\sqrt{2}}\,\mu \nonumber \\
g_{H,l} & = - \frac{\lambda_\tau}{\sqrt{2}}\,\mu &
g_{H,\tau} & = \frac{1}{2}\,{g^\prime}^2\,v_2 \;\;\;. & & 
\end{align}
If we perform the substitution $M^2_H \leftrightarrow M^2_h$ in 
equations~\eqref{eq:mh1}-~\eqref{eq:mh3} we obtain the second derivatives of the
mass of the heaviest scalar Higgs. Finally, for the staus we have
\begin{eqnarray}
M^2_{\tilde{\tau}_1,ll} & = & \frac{({g^\prime}^2 + g_2^2)\,\bigl(j_{\bar{H}}
\,-\,M^2_{\tilde{\tau}_1}\bigr) \,+\,(2 \lambda_\tau^2 - {g^\prime}^2)\,\bigl(
h_{\bar{H}}\,-\,M^2_{\tilde{\tau}_1}\bigr)}{M^2_{\tilde{\tau}_2}\,-\,
M^2_{\tilde{\tau}_1}} \;+\;\frac{A_{ll}(M^2_{\tilde{\tau}_1})}{\bar{
D}_{\tilde{\tau}}\bigl(M^2_{\tilde{\tau}_1},M^2_{\tilde{\tau}_2}\bigr)}\;+ 
\nonumber \\
 & & \frac{B_{ll}(M^2_{\tilde{\tau}_1})}{D_{\tilde{\tau}}\bigl(
M^2_{\tilde{\tau}_1},M^2_{\tilde{\tau}_2}\bigr)} \nonumber \\
M^2_{\tilde{\tau}_1,\tau\tau} & = & \frac{(2 \lambda_\tau^2 - {g^\prime}^2)\,
\bigl(j_{\bar{H}}\,-\,M^2_{\tilde{\tau}_1}\bigr) \,+\,4\,{g^\prime}^2\,\bigl(
h_{\bar{H}}\,-\,M^2_{\tilde{\tau}_1}\bigr)}{M^2_{\tilde{\tau}_2}\,-\,
M^2_{\tilde{\tau}_1}} \;+\;\frac{A_{\tau\tau}(M^2_{\tilde{\tau}_1})}{\bar{ 
D}_{\tilde{\tau}}\bigl(M^2_{\tilde{\tau}_1},M^2_{\tilde{\tau}_2}\bigr)}\;+ 
\nonumber \\
 & & \frac{B_{\tau\tau}(M^2_{\tilde{\tau}_1})}{D_{\tilde{\tau}}\bigl(
M^2_{\tilde{\tau}_1}, M^2_{\tilde{\tau}_2}\bigr)} \nonumber \\
M^2_{\tilde{\tau}_1,l\tau} & = & \frac{({g^\prime}^2 - 2 \lambda_\tau^2)\,
i_H}{M^2_{\tilde{\tau}_2}\,-\,M^2_{\tilde{\tau}_1}} \;+\;\frac{A_{l\tau}
(M^2_{\tilde{\tau}_1})}{\bar{D}_{\tilde{\tau}}\bigl(M^2_{\tilde{\tau}_1},
M^2_{\tilde{\tau}_2}\bigr)}\;+\;\frac{B_{l\tau}(M^2_{\tilde{\tau}_1})}{
D_{\tilde{\tau}}\bigl(M^2_{\tilde{\tau}_1},M^2_{\tilde{\tau}_2}\bigr)} \;\;\; ,
\end{eqnarray}
with
\begin{eqnarray}
\bar{D}_{\tilde{\tau}}\bigl(M^2_{\tilde{\tau}_1},M^2_{\tilde{\tau}_2}\bigr) &=& 
\bigl(M^2_{\tilde{\tau}_2}\,-\,M^2_{\tilde{\tau}_1}\bigr)\,\left[\bigl(
a_{\bar{H}}\,-\,M^2_{\tilde{\tau}_1}\bigr)\,\bigl(c_{\bar{H}}\,-\,
M^2_{\tilde{\tau}_1}\bigr)\,-\,b_{\bar{H}}^2 \right] \nonumber \\
D_{\tilde{\tau}}\bigl(M^2_{\tilde{\tau}_1},M^2_{\tilde{\tau}_2}\bigr) & = &
\bigl(M^2_{\tilde{\tau}_2}\,-\,M^2_{\tilde{\tau}_1}\bigr)\,\left[\bigl(
a_H\,-\,M^2_{\tilde{\tau}_1}\bigr)\,\bigl(c_H\,-\, M^2_{\tilde{\tau}_1}\bigr)\,-
\,b_H^2 \right] \;\;\; .
\end{eqnarray}
Again, the expressions for $M^2_{\tilde{\tau}_2}$ are obtained from these with
a simple replacement. Because supersymmetry is softly broken the supertrace
of the squared masses must be field-independent, so we can verify these formulae
by checking that $Str\, \partial^2 M^2/\partial x \partial y \, = \, 0$. 

\section{Numerical results and discussion}
\begin{figure}[htb]
\begin{center}
\epsfig{height=8cm,file=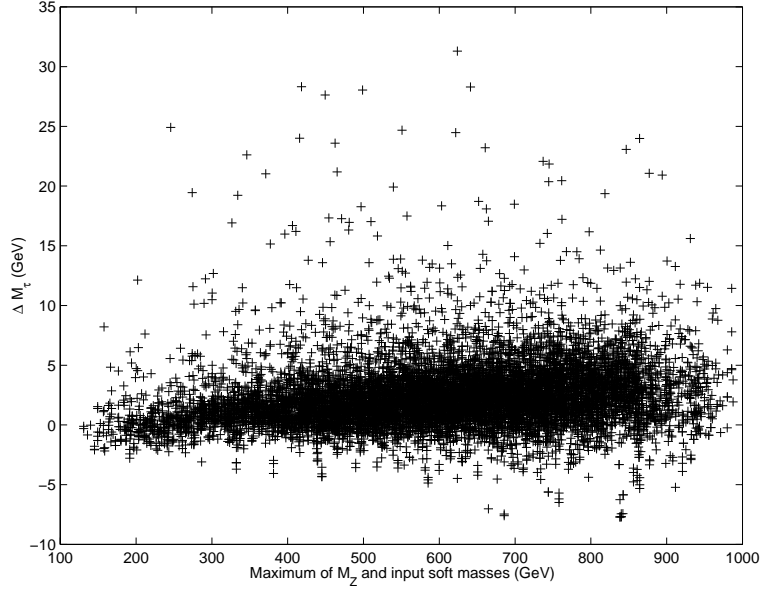}
\end{center}
\caption{The maximum $M$ of $M_Z$ and the input soft masses versus the mass 
difference between the one-loop and tree-level calculated lightest stau mass, 
$\Delta M_{\tilde{\tau}} \,=\, M_{\tilde{\tau}}^{1-loop}\,-\,
M_{\tilde{\tau}}^{tree}$.}
\label{fig:sca}
\end{figure}

\begin{figure}[htb]
\begin{center}
\epsfig{height=8cm,file=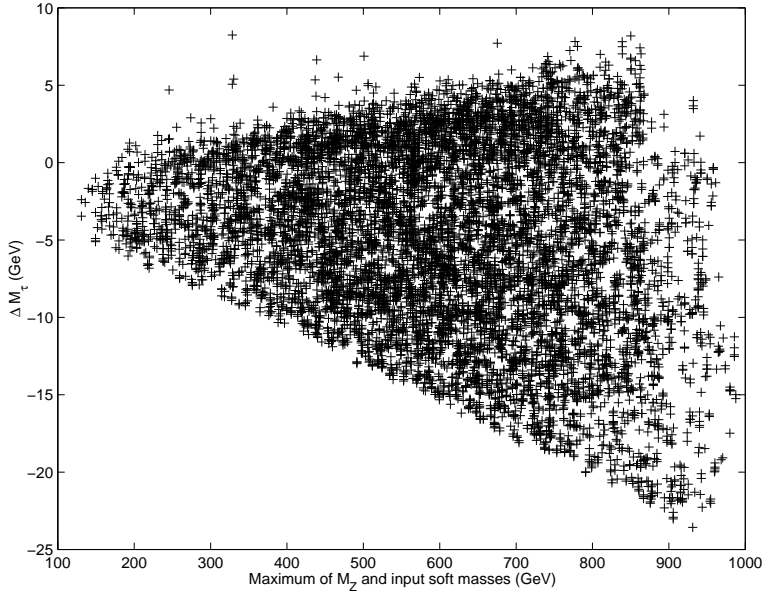}
\end{center}
\caption{The maximum $M$ of $M_Z$ and the input soft masses versus the mass 
difference between the one-loop and tree-level calculated heaviest stau mass.}
\label{fig:tau1}
\end{figure}

\begin{figure}[htb]
\begin{center}
\epsfig{height=8cm,file=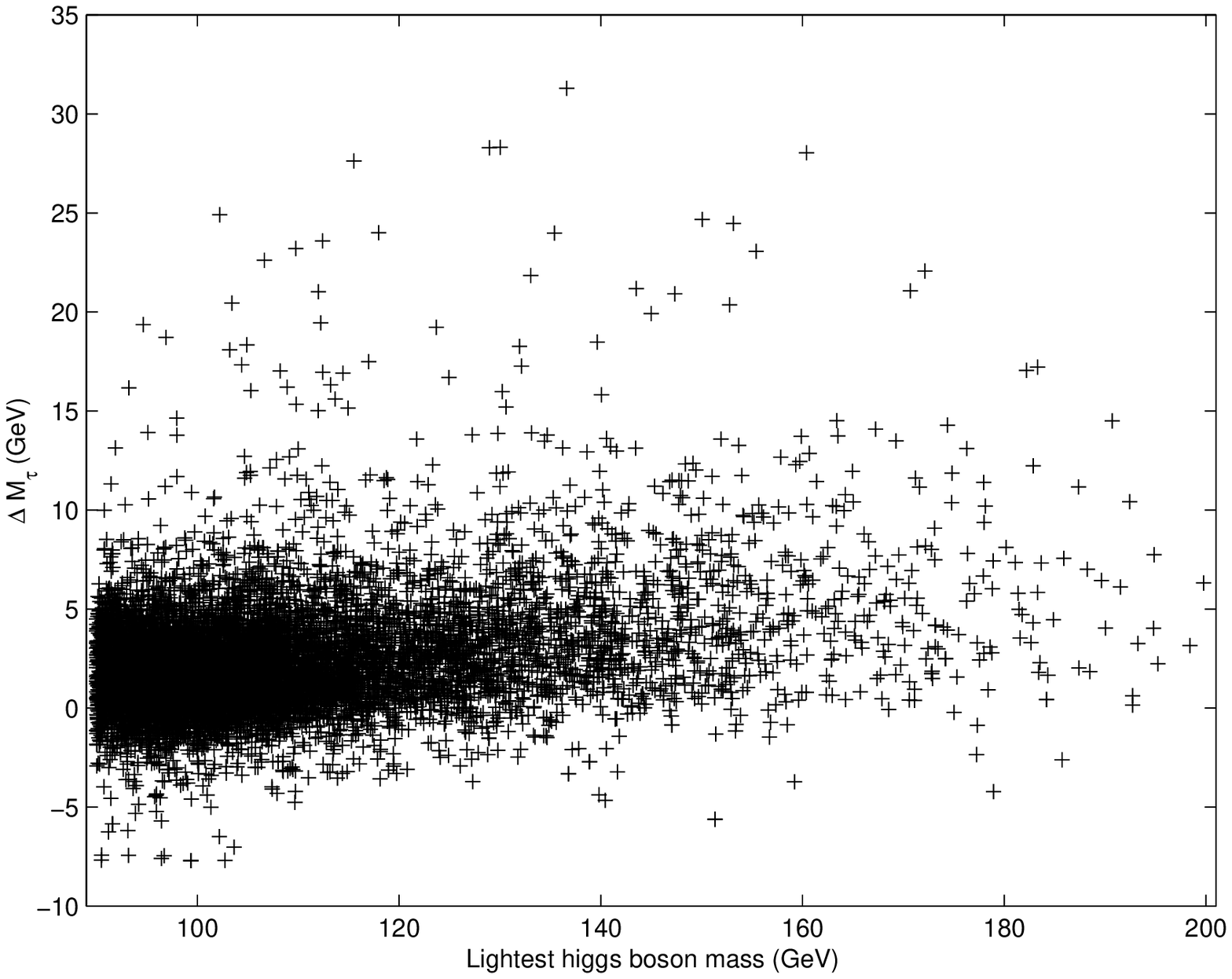}
\end{center}
\caption{Lightest CP-even higgs boson mass (one-loop) versus the mass difference
between the one-loop and tree-level calculated lightest stau mass.}
\label{fig:hig}
\end{figure}

We now apply our results for the one-loop stau masses to a vast MSSM parameter
space. In order to try to take into account the effects of the particles' mass 
thresholds in the renormalisation running of the theory's parameters, we follow
the procedure outlined in refs.~\cite{ros} and use as input parameters $\alpha_1
= 0.01667$, $\alpha_2 = 0.032$, $\alpha_S = 0.1$, $m_b = 2.95$ GeV, $m_\tau = 
1.75$ GeV and $m_t = 167.2$ GeV, at the scale $M_Z$. Accordingly we take the
DRED value for $v^2 = v_1^2 + v_2^2 = (250.75 \mbox{GeV})^2$, and use the 
supersymmetric two-loop $\beta$-functions to evolve all parameters between $M_Z$
and the gauge unification scale $M_U$, defined as the point where the couplings 
$\alpha_1$ and $\alpha_2$ meet. At $M_U$ we input the values of the soft 
parameters and determine $\mu$ and $B$  by minimising the one-loop MSSM 
potential at the scale $M$, defined as the maximum of $M_Z$ and the input scalar
and gaugino masses. For the soft parameters our strategy was to choose a random
value $M_G$, $m_G$ and $A_G$ and let the gaugino and scalar masses and $A$ 
parameters vary randomly within a 30\% interval of those central values, thus 
obtaining over 15000 points with input soft masses roughly in the interval
$[10, 1000]$ GeV and $-4 < A_G < 4$ TeV. Further, we have taken $2.5 \leq \tan
\beta \leq 6.5$ and considered both possible signs for the $\mu$ parameter. We 
then impose experimental bounds on the sparticles' masses from 
ref.~\cite{pdg}, except, obviously, the bounds on the stau masses, as we are 
interested in checking whether the one-loop contributions change their values 
considerably. As it turns out, for our choice of parameter space, after all 
other experimental cuts have been applied the remaining points (over 10000 of
them) correspond to stau masses above the current experimental bound (81 GeV) 
for all but a handful of points. The results of this ``scan" of the MSSM can be 
seen in figures~\eqref{fig:sca}-~\eqref{fig:hig}. In fig.~\eqref{fig:sca} we 
plot the mass difference between the one-loop and tree-level masses for the 
lightest stau, against the maximum $M$ of the input soft masses and $M_Z$ - 
$M$ is of the order of the largest masses present in $\Delta V_1$ and as such 
should constitute a good choice for renormalisation scale. Several observations 
about this plot: for $M$ smaller than about 200 GeV there is no substantial 
difference between the tree-level and one-loop results. For larger values, 
though, there are sizeable differences, usually smaller than 10 GeV. We observe 
that for the majority of points in the chosen parameter space the mass of the 
lightest stau increases. As expected, the one-loop contributions are small 
(typically less than 5\%) but we find they are not negligible. 
Figure~\eqref{fig:tau1} is the analogous of the previous one, but looking at the
one-loop/tree-level mass difference for the heaviest stau - we see the one-loop 
contributions tend to decrease the mass of the heavier stau, by as much as 
$\sim$ 20 GeV. Again, these contributions are only a few percent of the total 
mass (typically less than 8\%) but not at all insignificant. Now, since the 
lightest CP-even higgs boson is likely to be discovered before the staus, it is 
useful to look at the relationship between $\Delta M_{\tilde{\tau}}$ and 
$M_h$ (we use the full one-loop higgs mass). In fig.~\eqref{fig:hig} we plot 
these two quantities one against the other - despite the fact that the largest 
values of $\Delta M_{\tilde{\tau}}$ occur (naturally) for higher values of the 
input soft masses, they do not necessarily correspond to large values of $M_h$. 

In conclusion, we computed the one-loop contributions to the stau masses in the
effective potential approach and showed they are usually quite small, but can
nevertheless be sizeable. In particular, the lighter stau mass is shown to 
increase for the majority of input values we considered. If the staus are ever 
discovered and their mass measured accurately, these mass differences could be 
instrumental in narrowing the parameter space of the MSSM. Caution must be 
exercised in reading these results, though: the e.p.a., we repeat, is an 
approximation to the real mass. For instance, as discussed in ref.~\cite{brig}, 
the resulting mass has a small renormalisation scale dependence. We confirmed 
that fact by changing the value of $M$, doubling it or reducing it to 100 GeV, 
but the resulting changes in $\Delta M_{\tilde{\tau}}$ are indeed very small. We
also recall that the e.p.a. results for the Higgs sector were off by a few GeV, 
but again from ref.~\cite{brig} we observe the e.p.a tends to 
{\em underestimate} the real masses. As such, our conclusion regarding the 
increase of the lightest stau mass should hold. Of course, the validity of the 
e.pa. can only be established by performing the full diagrammatic calculation, 
and perhaps the fact the e.p.a. is predicting measurable differences for 
$M_{\tilde{\tau}}$ is sufficient reason to undertake it. We also observe that an
e.p.a. calculation in the stop sector should yield larger one-loop contributions
for the simple reason it produces second derivatives of the masses proportional 
to $\lambda_t^2$. This work is now under preparation.


\begin{thebibliography}{99}
\bibitem{ccb} P.M. Ferreira, {\em Phys. Lett.} {\bf B509} (2001) 120.
\bibitem{min} P.M. Ferreira, {\em Phys. Lett.} {\bf B512} (2001) 379.
\bibitem{erz} J. Ellis, G. Ridolfi and F. Zwirner, {\em Phys. Lett.} {\bf B262}
(1991) 477; {\bf B257} (1991) 83.

A. Brignole, J. Ellis, G. Ridolfi and F. Zwirner, {\em Phys. Lett.} {\bf B271}
(1991) 123.
\bibitem{brig} A. Brignole, {\em Phys. Lett.} {\bf B277} (1992) 313; {\bf B281}
(1992) 284.
\bibitem{ram} J.R. Espinosa and R. Zhang, {\em JHEP} {\bf 0003:026} (2000). 
\bibitem{bbo} V. Barger, M.S. Berger and P. Ohmann, {\em Phys. Rev.} {\bf D49}
(1994) 4908.
\bibitem{ros} L. Roszkowski, {\bf hep-ph/9509273}.

P. Langacker and N. polonski, {\em Phys. Rev.} {\bf D47} (1993) 4028.

M. Carena, S. Pokorski and C.E.M. Wagner, {\em Nucl. Phys.} {\bf B406} (1993) 
59.
\bibitem{pdg} D.E. Groom {\em et al}, {\em Eur. Phys. Jour.} {\bf C15} (2000) 1.
\bibitem{gam} G. Gamberini, G. Ridolfi and F. Zwirner, {\em Nucl. Phys.} {\bf
B331} (1990) 331.
\bibitem{rai} Y. Fujimoto, L. O'Raifeartaigh and G. Parravicini, {\em Nucl.
Phys.} {\bf B212} (1983) 268.
\end{thebibliography}
\end{document}